\documentclass[conference,10pt]{IEEEtran}
\usepackage{cite}

\usepackage{graphicx}
\usepackage{epstopdf}

\usepackage[cmex10]{amsmath}
\usepackage{amsthm}
\usepackage{amssymb}
\usepackage{cases}
\usepackage{bm}

\usepackage[caption=false,font=footnotesize]{subfig}

\usepackage{fixltx2e}

\usepackage{url}

\usepackage{algorithm}
\usepackage{algorithmic}
\begin{document}
\title{A Two-Step Learning and Interpolation Method for Location-Based Channel Database }

\author{\IEEEauthorblockN{Ruichen Deng\IEEEauthorrefmark{1}, Zhiyuan Jiang\IEEEauthorrefmark{1}, Sheng Zhou\IEEEauthorrefmark{1}, Shuguang Cui\IEEEauthorrefmark{2}, and Zhisheng Niu\IEEEauthorrefmark{1}}
        \IEEEauthorblockA{\IEEEauthorrefmark{1}Tsinghua National Laboratory for Information Science and Technology,\\
        	Department of Electronic Engineering, Tsinghua University, Beijing 100084, China\\
        	Email: drc13@mails.tsinghua.edu.cn, \{zhiyuan,sheng.zhou,niuzhs\}@tsinghua.edu.cn}

		\IEEEauthorblockA{\IEEEauthorrefmark{2} Department of Electrical and Computer Engineering, University of California, Davis, California 95616, USA\\     
		Email: sgcui@ucdavis.edu}}

\maketitle

\begin{abstract}
Timely and accurate knowledge of channel state information (CSI) is necessary to support scheduling operations at both physical and network layers. In order to support pilot-free channel estimation in cell sleeping scenarios, we propose to adopt a channel database that stores the CSI as a function of geographic locations. Such a channel database is generated from historical user records, which usually can not cover all the locations in the cell. Therefore, we develop a two-step interpolation method to infer the channels at the uncovered locations. The method firstly applies the K-nearest-neighbor method to form a coarse database and then refines it with a deep convolutional neural network. When applied to the channel data generated by ray tracing software, our method shows a great advantage in performance over the conventional interpolation methods.
 
\end{abstract}
\IEEEpeerreviewmaketitle

\section{Introduction}
In wireless communications, channels suffer from random variations due to multipath fading, shadowing, rain attenuation and other factors. Accordingly, accurate channel state information (CSI) could improve the performance of operations at both the physical and network layers, such as MIMO precoding and user association to base stations (BSs). The conventional method of channel acquisition is to use probing signals (pilots) known by both the transmitter and receiver. For example, the CSI-RS is established as a reference signal to obtain the channel state feedback of up to 8 transmit antennas in LTE Release 12\cite{Sesia2011LTE}. But pilot-based methods face a big challenge when applied in green networks, where the sleeping mechanism is usually adopted to save energy for BSs with light traffic load \cite{EDT_HCN}. To send pilots, sleeping BSs need to wake up periodically , which will dramatically decrease the performance of energy saving. 

Fortunately, the CSI can be acquired by exploiting its correlation with other entities. For the channel of a sleeping small BS in heterogeneous networks, Ref. \cite{CL} explores its correlation with the channel of the antenna array located at the macro BS. The relationship between the two channels is learned from training samples by a customized neural network (NN). On the other hand, the CSI is also strongly related to the user locations, which can be acquired by GPS and the trilateration method. In a stationary propagation environment, the CSI can be viewed as a function of the user location \cite{Locationaware,Locationaided}. Therefore, we can adopt a channel database that stores the CSI of different locations to support pilot-free channel estimation in the cell sleeping scenario.
Such a database is built based on the historical user records. Due to the nonuniform user distributions, the location entries of the database cannot be all covered. The CSI of uncovered locations need to be interpolated. 

Field tests have shown that there exist correlations among the channels at different locations, which typically decay exponentially with the distances between correlating parties \cite{sorensen1999slow,graziano1978propagation}. Based on the exponential decay model for the channel correlation and the well-known log-normal model for shadowing, the location-channel function can be treated as a Gaussian process and the MMSE estimator is applied to interpolate for the channel database construction \cite{Locationaware}. Although the Gaussian process model is easy for analysis, it is not accurate enough in practice since the model does not explore cell-specific channel correlations at different locations. Supervised learning is a data-driven technique that explores implicit correlations between two metrics.  A typical learning method for the irregular data distribution problem is the K-nearest-neighbor (KNN) \cite{KNN}, which predicts the channel as the average of its neighboring channels. The KNN method outperforms the Guassian process model based method in our simulations, since it explores the cell-specific characteristics from the historical data.

The combination weights in the averaging operation of the KNN method are not optimized, which makes the KNN interpolation results still rough. Therefore, we propose a two-step learning and interpolation method to further improve the performance. First, we apply the KNN method to interpolate for the channels at uncovered locations, which leads to a coarse channel database. Afterwards, a deep convolutional neural network (CNN) is used to refine the coarse database by exploring local channel patterns. We generate urban channel samples based on the ray tracing software to validate the performance. The proposed method is found to reduce the interpolation error by more than 50\% compared to the Guassian-process-based interpolation and the KNN interpolation. The proposed method can be easily extended to other propagation environments due to its model-free characteristic.


The rest of the paper is organized as follows. Section II introduce the location-based channel database and its matrix representation. Then the Gaussian process model for the channel database is illustrated in Section III. After that, we propose the two-step learning and interpolation method in Section IV. The simulation results are given in Section V and conclusions are drawn in Section VI.

\section{Location-Based Channel Databases}
\begin{figure}[!t]
	\centering
	\includegraphics[width=3.5in]{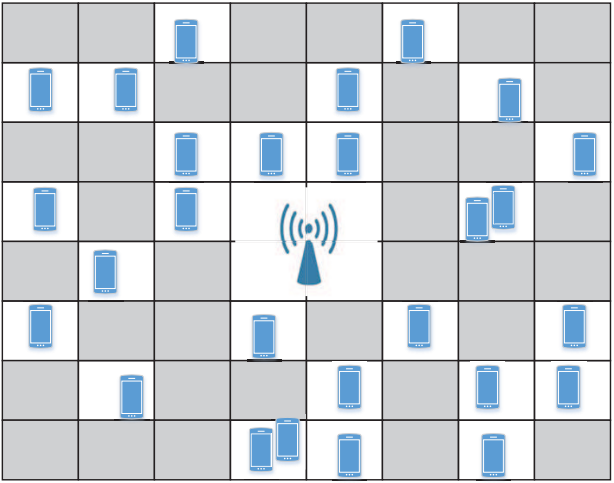}
	\caption{An illustration for the channel database. The cell is in a grid layout with the BS in the center and user samples in an irregular distribution. The invalid regions are colored in gray.}
	\label{fig:database}
\end{figure}
We consider a cell in a two-dimensional layout, where a location-based channel database of the BS stores the channel gains to different locations. Usually, the propagation environment is non-stationary and the channels are time-varying. As a result, the database should be renewed periodically to adapt to the changes. We consider one renewal process and assume the channels are approximately time-invariant. 

To renew such a database, an online data collecting approach is adopted, as shown in Fig. \ref{fig:database}. More specifically, the data samples were generated by the users who connected to the BS in the previous renew period (which was a subset of all users). The $n$-th user ($n=1,2,\cdots,N$) measured its channel from the BS, which is denoted as $y_{n}$, and sent it back to the BS along with its location coordinate $\bm{x}_n=(x_{n1},x_{n2})$. Compared to offline data collecting methods, this online approach can save the cost of channel measurement and well adapt to the changing propagation environment. But the online collected data is usually distributed irregularly over geographic locations due to the randomness of user distributions. 

The geographic region in the channel database is quantized in a grid layout to support fast inference.  More specifically, we use a matrix $\bm{D}$ to record the channel samples at different user locations. Let $x_{i,\text{min}}$ and $x_{i,\text{max}},~i=1,2$ be the lower and upper bounds of the $i$-th dimension coordinate, and $q$ be the quantization resolution. Then the channel record at location $(x_1,x_2)$ is mapped to the $\left(\text{R}\{\frac{x_1-x_{1,\text{min}}}{q}\}+1,\text{R}\{\frac{x_2-x_{2,\text{min}}}{q}\}+1\right)$-th entry of $\bm{D}$, where the operator $\text{R}\{\cdot\}$ rounds the value to the nearest integer. If there are more than one records mapped to the same entry, the value of the entry is set as the average of these records. If there are no records mapped to the entry, it is set to 0, indicating invalid data. We assume the set of valid data is $\mathcal{V}$ with cardinality $V$, and the valid entries are listed as  $\bm{v}_i=(v_{i1},v_{i2}),~i=1,2,\cdots,V$.

This mapping process actually divides the whole area into $H\times W$ disjoint square regions with size $q\times q$, where $H=\text{ROUND}\{\frac{x_{1,\text{max}}-x_{1,\text{min}}}{q}\}+1, W=\text{ROUND}\{\frac{x_{2,\text{max}}-x_{2,\text{min}}}{q}\}+1$. The channel of each region is represented by the corresponding entry of $\bm{D}$. By decreasing $q$, we can get larger $\bm{D}$ and $\bm{M}$, which implies a more precise description of the channel database. However, the ratio of valid entries in $\bm{D}$ also decreases given the same amount of historical data, making the training more difficult. Therefore, the proper choice of $q$ should reach a balance between training efficiency and quantization accuracy of the database.

Our goal is to complete the channel database matrix $\bm{D}$, i.e., to predict the invalid entries in $\bm{D}$ based on the valid entries. In other words, we develop the interpolation function $g$ that maps the invalid entry index $(i,j)$ to the channel:
\begin{equation}
g \colon (i,j) \mapsto \hat{y}_{ij},
\end{equation}
where $\hat{y}_{ij}$ is the predicted value of $y_{ij}$. 

\section{Model-based Channel Interpolation}
\subsection{Channel Model}
In the conventional channel database construction, the channel $y_{ij}$ is usually modeled as a Gaussian process \cite{Locationaware}. More specifically, according to the log-distance path loss model, the channel could be expressed as (in a dB scale)
\begin{equation}\label{pathloss}
	y_{ij} = G_0-10\eta\log_{10} L_{ij}+\psi_{ij}+\zeta_{ij},
\end{equation} 
where $G_0$ is a constant related to antenna gain, $\eta$ is the path loss exponent, $L_{ij}$ is the Euclidean distance between the transmitter and receiver, $\psi_{ij}$ represents the shadowing, and $\zeta_{ij}$ represents all other (non-shadowing) losses. The fading term $\zeta_{ij}$ is usually dominated by multipath fading, which is independent of the shadowing term $\psi_{ij}$ and has zero correlation over distances greater than a few
wavelengths. The total fading is modeled as a log-normal distribution with $\psi_{ij}+\zeta_{ij} \sim \mathcal{N}(0,\sigma_\psi^2+\sigma_\zeta^2)$ \cite{Agrawal09}, where $\sigma_\psi^2$ and $\sigma_\zeta^2$ are the variations of shading and non-shadow fading, respectively. The spatial covariance function of shadowing is given by \cite{gudmundson1991correlation}
\begin{equation}
	\mathbb{E}\{\psi_{ij}\psi_{lm}\} = \sigma_\psi^2 \exp\left(-\frac{d(ij,lm)}{d_0}\right),
\end{equation}
where $d({ij,lm})=\sqrt{(i-l)^2+(j-m)^2}$ is the distance between coordinates $(i,j)$ and $(l,m)$, and $d_0$ is the correlation distance.

We assume the channels at different locations follow a multivariant Gaussian distribution:
\begin{equation}
y_{11},y_{12},\cdots,y_{HW} \sim \mathcal{N}(\bm{u},\bm{C}),
\end{equation}
where $\bm{u}$ is the mean vector with the $((i-1)W+j)$-th element as $G_0-10\eta\log_{10} L_{ij}$, and $\bm{C}$ is the covariance matrix with the $\left((i-1)W+j,(l-1)W+m\right)$-th entry as
\begin{equation}
C_{(i-1)W+j,(l-1)W+m}=
\begin{cases}
\sigma_\psi^2+\sigma_\zeta^2&i=l,j=m,\\
\sigma_\psi^2 \exp\left(-\frac{d(ij,lm)}{d_0}\right)&else.
\end{cases}
\end{equation}.


Given an invalid entry $(i,j)$ of $\bm{D}$, the covariance vector between this entry and valid entries is calculated as 
\begin{equation}
\bm{a}_{ij}=[\sigma_\psi^2 \exp(-\frac{d({ij,v_{11} v_{12}})}{d_0}),\cdots,\sigma_\psi^2 \exp(-\frac{d({ij,v_{V1}v_{V2}})}{d_0})]^T.
\end{equation}
Therefore, the channel of this entry is estimated as \cite{kay1993fundamentals}
\begin{equation}\label{est}
\hat{y}_{ij} = \bm{a}_{ij}^T\bm{C}_{\mathcal{V}}^{-1}(\bm{y}_{\mathcal{V}}-\bm{u}_{\mathcal{V}})+u_{ij},
\end{equation}
where $\bm{C}_{\mathcal{V}}$, $\bm{y}_{\mathcal{V}}$ and $\bm{u}_{\mathcal{V}}$ are the covariance matrix, the channel vector and the mean vector of the valid entries. The minimum MSE is thus achieved:
\begin{equation}\label{est_MSE}
\epsilon_{uk} = \sigma_\psi^2+\sigma_\zeta^2-\bm{a}_{ij}^T\bm{C}_{\mathcal{V}}^{-1}\bm{a}_{ij}.
\end{equation}

The matrix inversion in the MMSE estimator involves huge computation costs when the number of samples is large. To reduce the complexity, we can approximate the MMSE estimator by using only neighborhood channels for prediction. Notice that the channel correlations are quite weak among locations with distances more than $3d_0$. Therefore we only choose the neighboring valid entries $\{(b_1,b_2)\in\mathcal{V}|~\vert b_1-i\vert\leq N_n,\vert b_2-j\vert\leq N_n\}$ to estimate the channel for the $(i,j)$-th entry, where $N_n$ is a predefined range of neighborhood. 
\subsection{Parameter Estimation}
The parameters in the Gaussian model is categorized into two categories:
\begin{enumerate}
	\item $G_0,\eta$, which are related to path loss;
	\item $d_0,\sigma_\psi^2+\sigma_\zeta^2$, which are related to inter-user correlations via shadowing;
\end{enumerate} 

These parameters are coupled together, which makes it difficult to derive a minimum variance unbiased estimator for them. The correlation of shadowing satisfies exponential delay law and becomes weak for far-away entries. Therefore we apply an approximate estimator, which first estimates the path loss parameters by treating all the samples as uncorrelated, and then estimates the correlation parameters of shadowing.
\subsubsection{Path Loss Parameter Estimation}
We rewrite the log-distance path loss model in (\ref{pathloss}) for all valid entries into the vector form:
\begin{equation}
\bm{y}_{\mathcal{V}} = \bm{L}_{\mathcal{V}}\begin{bmatrix}
G_0\\
\eta
\end{bmatrix}+\bm{\psi}_{\mathcal{V}}+\bm{\zeta}_{\mathcal{V}},
\end{equation}
where $\bm{L}_{\mathcal{V}}=[\bm{1}_{N\times 1}, -10\log_{10}\bm{l}_{\mathcal{V}}]$ with $\bm{l}_{\mathcal{V}}=[L_{v_{11} v_{12}},\cdots,L_{v_{V1} v_{V2}}]^T$. The estimation of $G_0$ and $\eta$ is similar to the linear model described in \cite{kay1993fundamentals}. We apply the least square estimator to obtain an estimation of the parameters given $\bm{y}_{\mathcal{V}}$:
\begin{equation}
\begin{bmatrix}
{G}_0^\star\\
{\eta}^\star
\end{bmatrix}=(\bm{L}_{\mathcal{V}}^T\bm{L}_{\mathcal{V}})^{-1}\bm{L}_{\mathcal{V}}^T\bm{y}_{\mathcal{V}}.
\end{equation}

\subsubsection{Estimating Correlation Parameters of Shadowing}
After the path loss parameters are estimated, the fading terms of each valid entry is obtained as
\begin{equation}
\hat{\psi}_{v_{i1}v_{i2}}+\hat{\zeta}_{v_{i1}v_{i2}}= y_{v_{i1}v_{i2}}- {G}_0^\star+10{\eta}^\star\log_{10}L_{v_{i1}v_{i2}}.
\end{equation}

The total variation $\sigma_\psi^2+\sigma_\zeta^2$ is estimated as the empirical variation of the estimated fading terms. By simple calculations, the MMSE estimator (\ref{est}) is observed to be independent of $\sigma_\psi$. Hence the estimation of $\sigma_\psi$ is unnecessary. We only need to estimate the correlation distance $d_0$. For each valid entry $(v_{i1},v_{i2})$, we derive the MMSE estimate $\hat{\psi}_{v_{i1}v_{i2}}$ from its neighboring valid entries with the total MSE as
\begin{equation}\label{totalMSE}
\epsilon_{T}=\frac{1}{V}\sum_{i=1}^{V}\left(\hat{\psi}_{v_{i1}v_{i2}}-\psi_{v_{i1}v_{i2}}\right)^2,
\end{equation}

which is a function of the correlation distance $d_0$. As the MMSE is hard to derive due to the matrix inversion in the estimator, We adopt a simple one-dimension search to find the optimal $d_0^\star$ that minimizes the function in (\ref{totalMSE}). 

The model-based channel database construction method introduced in this section
has various performance limitations. We next propose a more advanced method based on data-driven machine learning.

\section{Two-Step Learning and Interpolation for the Channel Database}

If we regard the channel entries as pixel values and view the whole database as an image, the channel database interpolation problem is similar to the image super resolution problem \cite{ImageSR}\cite{SRCNN}, which is to enlarge an image by interpolating pixels. But the valid "pixels" of the channel database are not in a regular layout, making the interpolation of the database more difficult. One the other hand, the mapping function $g$ from locations to channels is too complex to learn directly, which is observed in the simulation results. As a result, we propose a two-step interpolation scheme, which first form a coarse channel database by the KNN method and then refines it by the CNN with a mask.

\subsection{Step I: Forming a Coarse Channel Database by KNN}
The matrix $\bm{D}$ obtained by utilizing the historical channel records in Section I has many invalid entries, which will hinder the application of CNN. Since the channel correlations at different locations decay dramatically with their distances, the KNN method suits the problem due to its nature of neighborhood-based interpolation. 

Therefore, we first form a coarse channel database with the help of KNN, which does not need a training stage. In this step, for an invalid entry $(u_1, u_2)$, it finds the $K$-nearest neighbors $(b_{11} b_{12}),\cdots,(b_{K1},b_{K2})$ in the  valid user set $\mathcal{V}$, and estimates the channel for $(u_1, u_2)$ as the weighted mean of its neighbor channels. Typically there are two ways of weight assignment. One is uniform weight assignment, which allocates equal weights to all $K$ neighbors (uniform-weight):
\begin{equation}
y_{u_1u_2}^{\text{uni}}=\frac{1}{K}\sum_{i=1}^{K}y_{b_{i1}b_{i2}}.
\end{equation}
The other is distance-based-weight assignment, which allocates weights proportional to the inverse of the distance from the neighbors:
\begin{equation}
y_{u_1u_2}^{\text{dis}}=\frac{\sum_{i=1}^{K}\frac{1}{d(b_{i1}b_{i2},u_1 u_2)}y_{b_{i1}b_{i2}}}{\sum_{i=1}^{K}\frac{1}{d(b_{i1}b_{i2},u_1 u_2)}},
\end{equation}
where $d(A,B)$ denotes the Euclidean distance between the input vectors of $A$ and $B$.

We apply the KNN operator $\mathcal{K}$ to the channel matrix $\bm{D}$ and get a coarse interpolated channel matrix $\bm{E}$.
\begin{figure}[!t]
	\centering
	\includegraphics[width=3.5in]{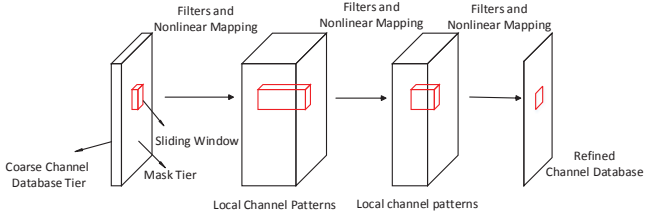}
	\caption{The structure of the convolutional neural network. The hidden layers apply filters and nonlinear mapping operator to the input to get the database features.}
	\label{fig:CNN}
\end{figure}
\subsection{Step II: Refining Channel Database by CNN}
The KNN operator can be viewed as an linear filter that averages the channels in the neighborhood of the location to be estimated. However, its average operation does not consider the local patterns of channel variations. To address this, a convolutional neural network $\mathcal{N}$ is used to transform the coarse channel matrix $\bm{E}$ to the refined channel matrix $\bm{F}$.
Such a CNN operation is also a neighborhood-based interpolation, which is similar to the filter operation with a sliding window. Unlike the conventional filters, the parameters of filters in the CNN are obtained by sample training and thus optimized. The preprocessing of the KNN interpolation in the previous step is necessary, since the CNN does not work well for incomplete input. 

For the CNN, we adopt the structure in Fig.\ref{fig:CNN}, which consists of $N_\text{CNN}$. The input $\bm{U}_1$ consists of two tiers\footnote{To prevent confusion, we use "tier" to represent a component of a data matrix instead of "channel" used in image processing.}, namely the coarse channel matrix $\bm{E}$ and the mask matrix $\bm{M}$. The mask matrix $\bm{M}$ has the same size as $\bm{D}$. It indicates whether the corresponding entry in $D$ is valid (1 for yes and 0 for no). The mask tier is added to indicate the validness of each entry in the data tier.

Each hidden layer $i~(i=1,2,\cdots,N_\text{CNN})$ conducts convolution operations to the input data $\bm{U}_i$ with $t_i$ tiers with size $C_i\times C_i$. The output has $T_{i+1}$ tiers representing the extracted channel patterns. The output of the $i$-th layer is the input of the $(i+1)$-th layer:
\begin{equation}
\bm{U}_{i+1}=\max (0,\bm{W}_i * \bm{U}_i+\bm{B}_i),
\end{equation}
where $\bm{W}_i$ is an $f_i\times f_i\times T_i\times T_{i+1}$ tensor representing $T_{i+1}$ linear filters and $\bm{B}_i$ is a $T_{i+1}$ dimensional vector representing the biases of all the tiers. 

More specifically, a sliding window with a size of $f_i\times f_i$ moves over the input with a stride of $s_i$. Inside the window, the inner product of the data and the coefficients of a linear filter is calculated as the entry of an output tier. After that, the rectified linear unit (ReLU) function is used as the activation function to increase nonlinearity \cite{relu}. A valid convolution (with no paddings on the edges) is applied. Therefore, each tier of the output has a size of $C_{i+1}\times C_{i+1}$ with
\begin{equation}
	C_{i+1}=\left\lfloor\frac{C_i-f_i}{s_i}\right\rfloor+1.
\end{equation}

The output $\bm{U}_{N_\text{CNN}}$ of the third hidden layer is the final output of the network, which has one tier ($T_{N_\text{CNN}}=1$).
  
\subsection{Training and Interpolation} 
We divide the valid data set randomly into two subset for training, namely the training set $\mathcal{T}$ and the labeling set $\mathcal{L}$. In the training process, we regard $\mathcal{T}$ as the new valid data set and use the channels in $\mathcal{T}$ to interpolate the channels in $\mathcal{L}$.

More specifically, we generate the channel matrix of the training set as $\bm{D}_\mathcal{T}$ and the mask matrix of the training set as $\bm{M}_\mathcal{T}$. As discussed before, the coarse interpolated channel matrix of training process is obtained by the KNN preprocessing:
\begin{equation}
\bm{E}_\mathcal{T}=\mathcal{K}(\bm{D}_\mathcal{T},\bm{M}_\mathcal{T}).
\end{equation}
The loss function $L(\mathcal{N})$ is the MSE between the network output and the ground-truth in the labeling set $\mathcal{L}$:
\begin{equation}
L(\mathcal{N}) =\lVert (\bm{M}_\mathcal{V}-\bm{M}_\mathcal{T})\odot(\mathcal{N}(\bm{E}_\mathcal{T},\bm{M}_\mathcal{T})-\bm{D}) \rVert^2,
\end{equation}
where $\odot$ is the elementwise multiplication. $\bm{M}_\mathcal{V}-\bm{M}_\mathcal{T}$ indicates the entry indexes of the labeling set $\mathcal{L}$. The loss function applies elementwise multiplication to exclude the MSE at the entries that are not in $\mathcal{L}$.

In the interpolation process, the channel and mask matrices of the whole data set are applied with the KNN operator:
\begin{equation}
\bm{E}_\mathcal{V}=\mathcal{K}(\bm{D}_\mathcal{V},\bm{M}_\mathcal{V}),
\end{equation}
and then the interpolation is the output of the neural network:
\begin{equation}
\bm{F}_\mathcal{V}=\mathcal{N}(\bm{E}_\mathcal{V},\bm{M}_\mathcal{V}).
\end{equation}
Since the channels of sampling locations are known, we replace the corresponding entries of $\bm{F}$ with the ground-truth data and obtain the final channel database in the matrix $\bm{G}$.
\begin{equation}
\bm{G}_\mathcal{V}=(1-\bm{M}_\mathcal{V})\odot\bm{F}_\mathcal{V}+\bm{D}_\mathcal{V}.
\end{equation}    

\section{Simulation Results}
\begin{figure}[!t]
	\centering
	\includegraphics[width=3.5in]{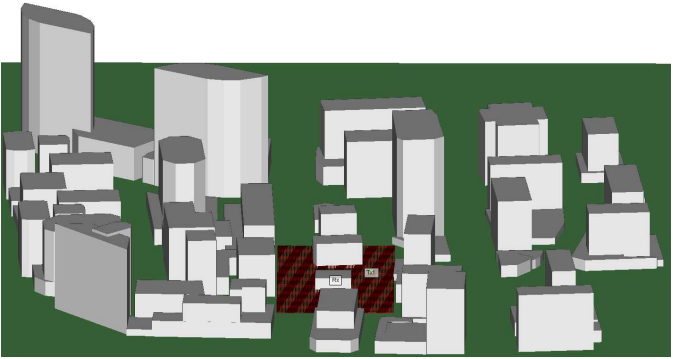}
	\caption{Simulation scenario. The coverage area of the BS is colored in red and the buildings are in white. The BS is located at Tx1 with 5 m high from the ground.}
	\label{fig:simu_setting}
\end{figure}
The software \emph{Wireless Incite} is used to generate ray tracing results for simulations \cite{WirelessInsite}. We simulate the propagation environment in Rosslyn, Virginia, which represents a typical heavily urbanized area, as is shown in Fig. \ref{fig:simu_setting}. The carrier frequency is set to 908 MHz and the bandwidth is 5 MHz. User samples are collected within a 150.0 m$ \times$ 84.0 m rectangle area. For simplicity, we consider a total of 50869 points in a grid layout with spacing 0.50 m and obtain the channels at these points by the ray tracing technique. The channels are all measured in a dB scale considering the wide dynamic range of the channel fading. Then we randomly choose half of the grid points (25434 points) as the valid sample set $\mathcal{V}$. The remaining samples form the testing set $\mathcal{T}$. We use channel information in the sample set  $\mathcal{C}$ to predict the channels in the testing set $\mathcal{T}$. The performance is measured by the average root mean square error (RMSE) between the interpolated channels and the ground-truth:
\begin{equation}
	e =\sqrt{\frac{1}{\vert \mathcal{T}\vert}\sum_{i\in\mathcal{T}}(g_i-y_i)^2},
\end{equation} 
where $g_i$ and $y_i$ are the interpolated result and the ground-truth channel of the $i$th sample in $\mathcal{T}$, respectively. We set the stride of the CNN to 1. The structure of CNN is represented by the symbol $f_1-f_2-\cdots-f_{N_\text{CNN}}~(T_1 - T_2 -\cdots T_{N_\text{CNN}})$, where $f_i$ and $T_i,~i=1,2,\cdots,N_\text{CNN}$ are the size and number of filters in the $i$-th layer, respectively. For example, the 9-1-5 (64-32-1) structure consists of 3 layers: the first layer has 64 filters with size $9\times9$, the second layer has 32 filters with size $1\times 1$, and the third layers has one filter with size $5\times5$.

\begin{figure}[!t]
	\centering
	\includegraphics[width=3.5in]{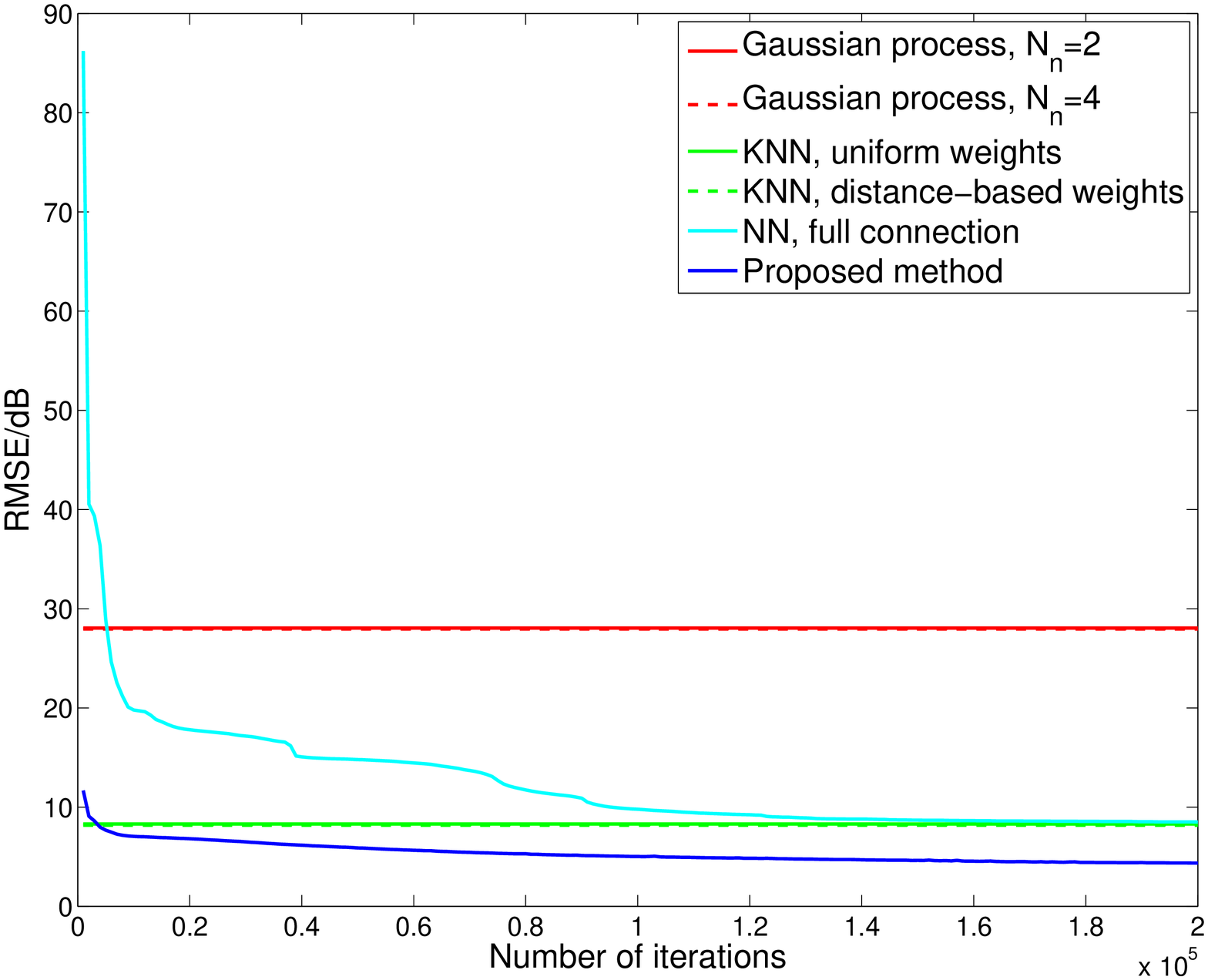}
	\caption{The performance of different methods. The number of neighbors in the KNN methods is 5. The CNN has a structure of 9-1-5 (64-32-1)}
	\label{fig:AlgCompare}
\end{figure}

First, we compare the performance of different algorithms in Fig. \ref{fig:AlgCompare}. The estimated pathloss parameters of the Gaussian process discussed in Section III are $ {G}_0^\star=3.26,~{\eta}^\star=-21.16~ \text{dB}$. We consider two choices for the predefined range of neighborhood: $N_n=2$ and $N_n=4$, which covers a square area of size $5\times5$ and $9\times9$, respectively. The estimated correlation distances of the two Gaussian processes both are 0.50 m. The RMSE of their interpolations are very close and much higher than the other methods. We use the channel average of $3$ neighbors for interpolation in the KNN method. The distance-based-weight KNN performs slightly better than the uniform-weight one. We also consider a full connected neural network structure (NN, full connection), which treats location coordinates as input and channels as output and directly learns the mapping $g$ between them from historical samples. After tuning paramters, the network adopt two hidden layers both with 10 nodes. The performance of this structure does not surpass the KNN method. This is due to the global mapping $g$ is too complex for the full connected network to learn. 

In the proposed method, we use the KNN with distance-based weights and the CNN with the 9-1-5 (64-32-1) structure. We adopt the Adam algorithm for each iteration of optimization \cite{DL}. After less than 5000 iterations, the network already outperforms the KNN method. After 200,000 iterations, the network reduces the RMSE from  8.20 dB of the KNN to  4.35 dB. The reduction is about 47\%.

\begin{figure}[!t]
	\centering
	\includegraphics[width=3.5in]{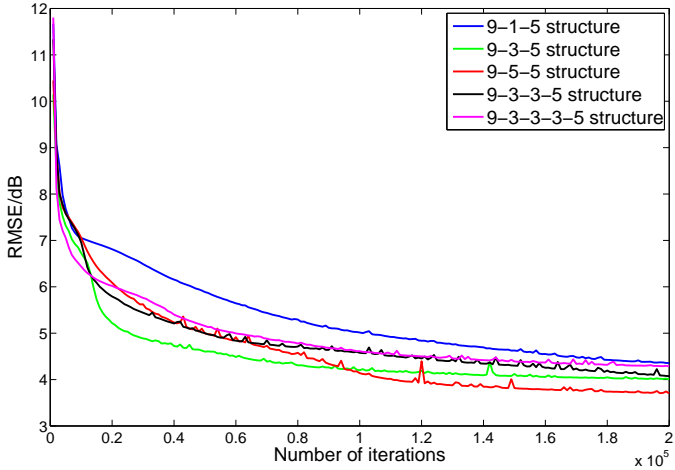}
	\caption{RMSE VS. filter size and number of layers.}
	\label{fig:RMSE_structure}
\end{figure}
\begin{figure}[!t]
	\centering
	\includegraphics[width=3.5in]{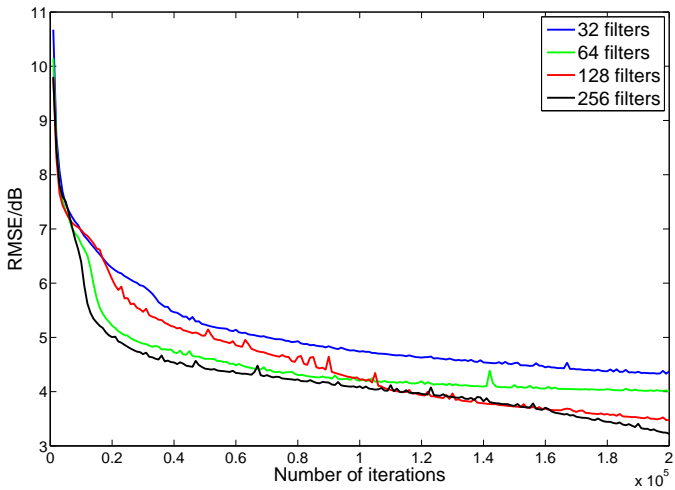}
	\caption{RMSE v.s. number of filters in the first layer. The filter sizes of the layers in the CNN are 9-3-5.}
	\label{fig:RMSE_filters}
\end{figure}
The performance of the CNN can be improved by tuning the hyperparameters. Fig. \ref{fig:RMSE_structure} shows the impact of filter size and number of layers on the performance. On one hand, increasing the filter size of the hidden layer can decrease the interpolation error. From the 9-1-5 structure to the 9-5-5 structure, the RMSE decreases by about $0.65$ dB and is reduced by about $55\%$ compared to the KNN interpolation. This is due to the fact that the representing ability of the network increases with the filter size. One the other hand, however, going deeper does not necessarily improve the performance. From the 9-3-5 structure to the 9-3-3-5 structure and the 9-3-3-3-5 structure, the RMSE gets higher rather than lower. A possible explanation for this result is the increased training difficulty with more layers. Since the network has no pooling or full connection layers, it is easily stuck in a bad local minimum. Similar results are observed in \cite{SRCNN}. Another factor affecting the performance is the number of filters, which is shown in Fig. \ref{fig:RMSE_filters}. Increasing the number of filters in the first layer from 32 to 256 can reduce the RMSE by about $1.1$ dB. The runtime of different network structures (normalized by the runtime of the 9-1-5 (64-32-1) structure) is presented in Table \ref{Tab:runtime}. It costs about $1024 s$ to finish 200000 training iterations for the 9-1-5 (64-32-1) structure on our platform. The training time for the network with larger filter sizes, more filters and deeper structures is obviously higher.

\begin{table}[!hbp]
	\renewcommand{\arraystretch}{1.3}
	\centering
	\caption{Runtime to train different network structures.}\
	\label{Tab:runtime}
	\begin{tabular}{c|c}
		\hline
		Network structure & Normalized runtime  \\
		\hline \hline
		9-1-5 (64-32-1) & 1.00\\
		\hline
		9-3-5 (32-32-1) & 1.09\\
		\hline
		9-3-5 (64-32-1) & 1.30\\
		\hline
		9-3-5 (128-32-1) & 1.68\\
		\hline
		9-3-5 (256-32-1) & 2.44\\
		\hline
		9-5-5 (64-32-1) & 1.55\\
		\hline
		9-3-3-5 (64-32-16-1) & 1.55\\
		\hline
		9-3-3-3-5 (64-32-16-16-1) & 1.93\\
		\hline
	\end{tabular}
\end{table}


\section{Conclusion}\label{sec:conclusion}
In this work, we considered the construction of channel databases and proposed a two-step learning and interpolation method to estimate the missing entries in the database. A coarse database was first built by the KNN method and then refined by the CNN. In the urban channels generated by the ray tracing software, we showed that the RMSE of KNN interpolation is much lower (about 20 dB less) than the Guassian process based approach. The proposed two-step method can further reduce the interpolation errors by more than 50\%. By testing different CNN architectures in our method, we find that the benefit of increasing network layers is much less than that of increasing the number or the size of filters.

\bibliographystyle{IEEEtran}
\bibliography{myref}

\begin{thebibliography}{10}
\providecommand{\url}[1]{#1}
\csname url@samestyle\endcsname
\providecommand{\newblock}{\relax}
\providecommand{\bibinfo}[2]{#2}
\providecommand{\BIBentrySTDinterwordspacing}{\spaceskip=0pt\relax}
\providecommand{\BIBentryALTinterwordstretchfactor}{4}
\providecommand{\BIBentryALTinterwordspacing}{\spaceskip=\fontdimen2\font plus
\BIBentryALTinterwordstretchfactor\fontdimen3\font minus
  \fontdimen4\font\relax}
\providecommand{\BIBforeignlanguage}[2]{{%
\expandafter\ifx\csname l@#1\endcsname\relax
\typeout{** WARNING: IEEEtran.bst: No hyphenation pattern has been}%
\typeout{** loaded for the language `#1'. Using the pattern for}%
\typeout{** the default language instead.}%
\else
\language=\csname l@#1\endcsname
\fi
#2}}
\providecommand{\BIBdecl}{\relax}
\BIBdecl

\bibitem{Sesia2011LTE}
S.~Sesia, M.~Baker, and I.~Toufik, \emph{{LTE}-the {UMTS} {Long} {Term}
  {Evolution}: {From Theory to Practice}}.\hskip 1em plus 0.5em minus
  0.4em\relax John Wiley \& Sons, Hoboken, NJ, Oct. 2011.

\bibitem{EDT_HCN}
Z.~Niu, X.~Guo, S.~Zhou, and P.~R. Kumar, ``Characterizing energy-delay
  tradeoff in hyper-cellular networks with base station sleeping control,''
  \emph{IEEE Journal on Selected Areas in Communications}, vol.~33, no.~4, pp.
  641--650, April 2015.

\bibitem{CL}
J.~Liu, R.~Deng, S.~Zhou, and Z.~Niu, ``Seeing the unobservable: Channel
  learning for wireless communication networks,'' in \emph{IEEE Global
  Communications Conference (GLOBECOM)}, Dec.. 2015, pp. 1--6.

\bibitem{Locationaware}
R.~Di~Taranto, S.~Muppirisetty, R.~Raulefs, D.~Slock, T.~Svensson, and
  H.~Wymeersch, ``Location-aware communications for {5G} networks: How location
  information can improve scalability, latency, and robustness of {5G},''
  \emph{IEEE Signal Processing Magazine}, vol.~31, no.~6, pp. 102--112, Nov.
  2014.

\bibitem{Locationaided}
D.~Slock, ``Location aided wireless communications,'' in \emph{International
  Symposium on Communications Control and Signal Processing (ISCCSP)}, May
  2012, pp. 1--6.

\bibitem{sorensen1999slow}
T.~B. Sorensen, ``Slow fading cross-correlation against azimuth separation of
  base stations,'' \emph{Electronics Letters}, vol.~35, no.~2, pp. 127--129,
  Jan. 1999.

\bibitem{graziano1978propagation}
V.~Graziano, ``Propagation correlations at 900 mhz,'' \emph{IEEE Transactions
  on Vehicular Technology}, vol.~27, no.~4, pp. 182--189, Nov. 1978.

\bibitem{KNN}
T.~Cover and P.~Hart, ``Nearest neighbor pattern classification,'' \emph{IEEE
  Transactions on Information Theory}, vol.~13, no.~1, pp. 21--27, January
  1967.

\bibitem{Agrawal09}
P.~Agrawal and N.~Patwari, ``Correlated link shadow fading in multi-hop
  wireless networks,'' \emph{IEEE Transactions on Wireless Communications},
  vol.~8, no.~8, pp. 4024--4036, Aug. 2009.

\bibitem{gudmundson1991correlation}
M.~Gudmundson, ``Correlation model for shadow fading in mobile radio systems,''
  \emph{Electronics Letters}, vol.~27, no.~23, pp. 2145--2146, Nov. 1991.

\bibitem{kay1993fundamentals}
S.~M. Kay, \emph{{Fundamentals of Statistical Signal Processing, Volume I:
  Estimation Theory}}.\hskip 1em plus 0.5em minus 0.4em\relax Prentice Hall,
  April 1993.

\bibitem{ImageSR}
J.~Yang, J.~Wright, T.~S. Huang, and Y.~Ma, ``Image super-resolution via sparse
  representation,'' \emph{IEEE Transactions on Image Processing}, vol.~19,
  no.~11, pp. 2861--2873, Nov. 2010.

\bibitem{SRCNN}
C.~Dong, C.~C. Loy, K.~He, and X.~Tang, ``Image super-resolution using deep
  convolutional networks,'' \emph{IEEE Transactions on Pattern Analysis and
  Machine Intelligence}, vol.~38, no.~2, pp. 295--307, Feb. 2016.

\bibitem{relu}
X.~Glorot, A.~Bordes, and Y.~Bengio, ``Deep sparse rectifier neural networks,''
  in \emph{Proceedings of the Fourteenth International Conference on Artificial
  Intelligence and Statistics}, June 2011, pp. 315--323.

\bibitem{WirelessInsite}
P.~Medeđović, M.~Veletić, and .~Blagojević, ``Wireless insite software
  verification via analysis and comparison of simulation and measurement
  results,'' in \emph{Proceedings of the 35th International Convention MIPRO},
  May 2012, pp. 776--781.

\bibitem{DL}
I.~Goodfellow, Y.~Bengio, A.~Courville, and Y.~Bengio, \emph{{Deep
  Learning}}.\hskip 1em plus 0.5em minus 0.4em\relax MIT Press, Cambridge,
  2016, vol.~1.

\end{thebibliography}

\end{document}